# Information Exchange on an Academic Social Networking Site: A Multi-discipline Comparison on ResearchGate Q&A[1]

Wei Jeng

School of Information Sciences, University of Pittsburgh, 135 North Bellefield Avenue, Pittsburgh, PA, USA

Email: wej9@pitt.edu

Spencer DesAutels[&]

Eskind Biomedical Library, Vanderbilt University Medical Center, Nashville, TN, USA

Email: spencer.j.goodwin@gmail.com

Daqing He

School of Information Sciences, University of Pittsburgh, 135 North Bellefield Avenue, Pittsburgh, PA, USA

Email: dah44@pitt.edu

Lei Li[&]

Department of Information Management, Nanjing University of Science and Technology, Nanjing, China

Email: lileiwelldone@gmail.com

[&]This work was done while these authors were at the School of Information Sciences, University of Pittsburgh.

## ABSTRACT

The increasing popularity of academic social networking sites (ASNSs) requires studies on the usage of ASNSs among scholars, and evaluations of the effectiveness of these ASNSs. However, it is unclear whether current ASNSs have fulfilled their design goal, as scholars' actual online interactions on these platforms remain unexplored. To fill the gap, this paper presents a study based on data collected from ResearchGate. Adopting a mixed-method design by conducting qualitative content analysis and statistical analysis on 1128 posts collected from ResearchGate Q&A, we examine how scholars exchange information and resources, and how their practices vary across three distinct disciplines: Library and Information Services, History of Art, and Astrophysics.

Our results show that the effect of a questioner's intention (i.e., seeking information or discussion) is greater than disciplinary factors in some circumstances. Across the three disciplines, responses to questions provide various resources, including experts' contact details, citations, links to Wikipedia, images, etc. We further discuss several implications of the understanding of scholarly information exchange and the design of better academic social networking interfaces, which should aim to stimulate scholarly interactions by minimizing confusion, improving the clarity of questions, and promoting scholarly content management.

## INTRODUCTION

Scholarly information exchange is tightly connected with information and communication technologies (Fry & Talja, 2007). With the help of the massive, instant, and dynamic social web infrastructure, it is possible for various scholarly activities to be conducted entirely online. Along with the popular generic social network services, we also see rapid growth of specialized social platforms that can "help scholars to build their professional networks with other researchers and facilitate their various activities when conducting research" (Jeng, He, & Jiang, 2015, p.890). In this paper, we refer to these social platforms as *academic social networking sites* (ASNS). Compared to a general SNS, an ASNS usually offers more specific features targeting academics (e.g., public profiles with research-oriented properties). Well-known examples of ASNSs include Academia.edu, Mendeley, and ResearchGate.

Scholars in Human Information Behavior (HIB) point out that peoples' information behaviors can be affected by various contextual factors such as discipline, occupation, tasks, and academic background (Case, 2012). Therefore, studying scholarly information exchange on ASNSs should consider discipline-specific characteristics, which are important contextual factors. At the same time, discipline independence is critical to discovering similarities across disciplines. ASNSs are ready for cross-discipline studies (Jiang, Ni, He, & Jeng, 2013; Oh & Jeng, 2011), since question-answering and small group discussions have been widely implemented in ASNSs across many disciplines.

Consequently, the goal of our study is to examine scholars' information exchange in the form of question-answering and small group discussions on an ASNS, and we conducted the study across three different disciplines: humanities (History of Art), social science (Library and Information Services), and natural science (Astrophysics). To enable deeper analysis of scholars' behaviors, we adopted a mixed-method design and selected ResearchGate's Q&A as the platform for our study.

ResearchGate (http://www.researchgate.net) is one of the most well-known ASNSs that supports various scholarly activities (Haustein et al., 2014). Because it requires all users to register with a valid email from an academic institution and to use their real names for posting content, ResearchGate may help users maintain high academic standards in their online behaviors, which is helpful in our study of their scholarly information exchange.

As with previous literature (Bowler et al., 2012; Savolainen, 2012), we recognize that a user on a social question and answering site (hereafter: social Q&A) might not always look for factual information, but may also engage in opinion sharing, emotional support, or advice seeking without aiming for a "right" answer. Therefore, in this study, we use the terms "Q&A discussion", "topic thread" and "Q&A thread" interchangeably, as they all indicate a thread that contains an initial post and responses, no matter whether the questioner is seeking a single answer or a discussion. The term "question initiator" or "questioner" is used for the user creating the first post, whether it is a question or not, and a "respondent" or "answerer" is any user replying to this initial post.

Specifically, we explore the following two research questions in this paper:

- RQ1: What kinds of questions do scholars ask on ResearchGate Q&A in three different disciplines? What are the characteristics of these questions?
- RQ2: How do other scholars respond to posted questions? What are the characteristics of the responses? What are the resources they provide to their peer users?

Under these research questions, we seek to explore the types of questions raised and discussed by scholars on the site (i.e., information seeking) and the characteristics of other scholars' responses (i.e., information providing). From a perspective of research contribution to human information behavior, we try to connect what we learn in RQ1 and RQ2 to digital systems design.

ASNSs provide us platforms for examining academic users' online activities (Jeng et al., 2015; Thelwall & Kousha, 2014). However, despite several studies on scholarly information exchange using digital resources (Pilerot, 2012; Talja, 2002), there is a significant opportunity to study scholars' exchange on these newer, social platforms. In the following sub-section, we review literature related to online information exchange, academic social networks, and social Q&A. We then position our work within this relevant framework at the end of this section.

## LITERATURE REVIEW

### Online Information Exchange among Scholars

The concept of "information exchange" remains open and ambiguous. Researchers often use different terms, such as "sharing", "transfer", "giving", or "providing" to represent information sharing activities (Fidel, 2012; Pilerot, 2012). In general, a definition for "information exchange" or "information sharing" can be understood as "the flow of information or knowledge transfer," where previous related work on information sharing or information exchange generally focuses on the "identification of common interests, beliefs, and norms; on the flow and transfer of information; or on co-existence and material conditions characterizing the site where sharing takes place" (Pilerot, 2015).

Information exchange among scholars or experts can be affected by several factors, including "information type and distribution, task features, group structure and composition, temporal features, member characteristics, discussion procedures, and communication technology" (Wittenbaum, Hollingshead, & Botero, 2004). In the current web setting, web-based discussion groups, forums and mailing lists are the preferred channels for scholars' information exchange on a daily basis (Pilerot & Limberg, 2011). However, when scholars from multiple disciplines were interviewed about how they used networked resources as tools for their scholarly information communication (Fry & Talja, 2007; Talja, Vakkari, Fry, & Wouters, 2007), their replies reveal that scholars' usage patterns of digital resources such as mailing lists and personal homepages vary greatly, suggesting that gateways or types of information channel play different roles "in the shaping of scholarly communication in the digital environment"(Fry & Talja, 2007).

Information exchange among peers exhibits complex patterns. For example, Liu and Tsai analyzed 14 small groups on an online class discussion forum, and observed several communication patterns among student groups, including centralized, distributed, group development impediment, ability impediment, and partial knowledge exchange (Liu & Tsai, 2008). Pena-Shaff and Nicholls also studied students' messages on a computer bulletin board system (BBS), and found that online discussions, compared to face-to-face group discussions, created much more complex interactions, for the online discussion had a much longer duration and could have a greater chance of "forming a larger discussion cluster of more than 10 participants" (Pena-Shaff & Nicholls, 2004).

The quality of information exchange in a group setting can be affected by some factors whereas not by others. For example, Liu and Tsai (2008) did not find a difference in quality of the group work between a central figure presiding over discussion and decentralized exchange patterns. However, they did find that groups that had trouble developing a group identity received lower scores than any other patterns. Their results also suggest that strategies related to group development and sociability were very critical for enabling effective information exchange during teamwork.

**Studies on Academic Social Networks**

As relatively new social platforms that are still refining their functionalities, ASNSs attract many studies exploring their user populations and usage characteristics. For example, Jeng, He, and Jiang (2015) found that the majority of Mendeley users were junior researchers. Thelwall and Kousha (2015) showed that ResearchGate is being used at academic institutions around the world with high impact universities having a higher aggregate score of their members on ResearchGate. The highest adoption of the ResearchGate platform was in the United States, whereas it was distinctly lacking members from Chinese institutions.

Another research focus is on users' behaviors on ASNSs. Mendeley users were mainly motivated by a need to seek information related to their research community, and were primarily focused on the features directly related to their research work, rather than "meeting more peers" or "expanding the professional network" (Jeng et al., 2015). Academia.edu users followed the trends of scholarly communication with faculty receiving more profile views than students, but did not conform to general social network norms, because female users were not more popular and influential than males (Thelwall & Kousha, 2014).

## Social Question and Answering Sites

According to Harper and Raban, a question and answering site is "designed to allow people to ask and respond to questions on a broad range of topics" (Harper & Raban, 2008). Social Q&A sites have adopted the Web 2.0 model with user-generated and user-rated content (Gazan, 2011).

Recent studies on question and answering sites can be broken into two categories: content based and user based (Shah, Oh, & Oh, 2009). Content-based studies usually present a holistic overview of question types or answer characteristics on social Q&A sites. For example, scholars have investigated four different online Q&A websites- "Yahoo! Answers, a community-based Q&A model; WikiAnswers, a collaborative Q&A model; the Internet Public Library (IPL), an expert-based Q&A model; and Twitter" (Choi, Kitzie, & Shah, 2012). The study suggests users are more likely to post opinion-seeking questions to Yahoo Answers.

User-oriented research focuses more on user participations and motivations. For example, Oh (2012) found out that answerers on Yahoo Answers are more likely to contribute out of a sense of altruism and for self-enjoyment. Researchers also investigated the motivations behind askers on Yahoo Answers. Choi, Kitzie, and Shah (2014) found that the most common motivation for askers is to learn, as users can gain knowledge themselves through acquiring information. Consistant with Oh (2012)'s results, the second most common motivation for askers was "Having fun asking a question on Yahoo! Answers."

Given that current studies have been carried out on information exchange, communication patterns, ASNSs, and social Q&A, the literature still lacks a conclusive understanding of how scholars exchange information on a dynamic, social platform such as ResearchGate. Through this study, we hope to help shed light on information exchange and communication among scholars through ASNSs.

The uniqueness and innovation of our study comes from two important aspects: *scholars*, and *content* generated by those scholars, both of which are the focus of scholarly information exchange on an academic social network site. Firstly, scholars may behave seriously and responsibly in the information exchange, even in an online environment, particularly under their real names like in ResearchGate. To these online scholars, ASNSs can be an online extension of their traditional academic network, in which they have a professional reputation and career development to maintain. We are curious about whether or not this would make scholars behave differently than users of generic social platforms, let alone anonymous users. The target audience, unlike those in Yahoo! Answers, are in some respects peers who most likely work in higher education, research institutions, or engage in professional work. Secondly, the online content, even though it is generated on ASNSs, can still contain academic jargon, terminology, equations, and theories. More importantly, the content may involves scholarly debates, where there could be no right or wrong answer. Therefore, our study draws some input from social Q&A literature but is more heavily indebted to scholarly communication literature.

# METHODOLOGY

## Study Site: ResearchGate

With seven million academic users as of 2015, ResearchGate is one of the most well-known academic social network sites. As an ASNS, it aims to help scholars build their professional profiles, share publications, and ask questions to their peers (Thelwall & Kousha, 2015). For example, ResearchGate enables scholars to upload their publications and build a personal profile regarding their research interests, affiliations, awards and other recognitions. It also provides some simple altmetric measures to each user, such as number of profile views and number of publication downloads by the scholarly community. Further, ResearchGate has a Q&A platform for scholars from all disciplines to discuss, interact, and find answers under various research topics. A topic on this platform is associated with a category or a tag to indicate the domain of the question, and it can be "followed" by a scholar so that notifications about all related activities are sent to the scholar.

Because of these features, we chose ResearchGate Q&A as our research site for investigating scholars' interaction and communication on ASNSs. Figure 1 shows a sample question – "What is the most meaningful three-dimensional art: sculptures, mobiles, or architecture?" – posted by a ResearchGate user. Other scholars provided answers to the question, followed this question, or used "upvote" or "downvote" to rate both the question and the answers.

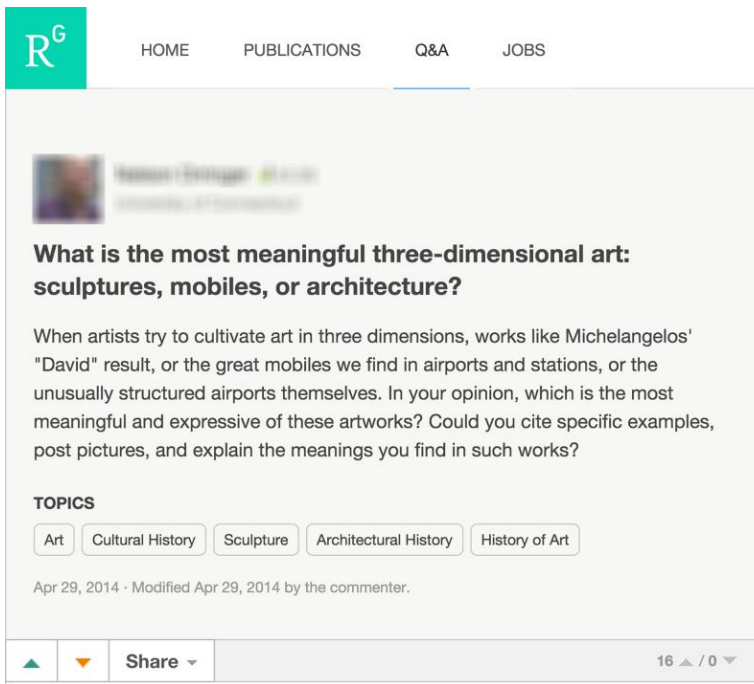

*Figure 1. A question thread on ResearchGate Q&A*

## Data Collection

We adopted a mixed-method design, which involves content analysis and a follow-up statistical analysis. Our research questions involve analyzing user-generated posts on ResearchGate Q&A. We also decided that our method should be manual content analysis because the types of analyses we want to perform on the questions and responses cannot be obtained through either quantitative methods or automatic methods. However, in order to compare the characteristics that we found among different disciplines, it was necessary to perform nonparametric statistical analysis to determine if the observation was statistically significant or just by chance.

Because scholarly information behaviors on ASNSs are relatively unknown in the literature, we decided to start with one discipline for developing our coding scheme, then expand the scheme to other disciplines. We considered three rationales in selecting our first discipline. First, the sample size should be manageable for manual coding, and preferably be comparable to previous non-ASNS studies in the literature (Liu & Tsai, 2008; Pena-Shaff & Nicholls, 2004). Second, in order to capture content-rich information and diverse patterns of conversations on ASNSs, the selected samples should contain large threads with many posts as well as small threads with few posts. Third, since we the authors will conduct the content analysis to develop the scheme, we wanted a certain level of domain knowledge in case such familiarity was needed in content analysis. Considering all these rationales, we chose the ResearchGate Q&A category "Library Information Services" as our first sample discipline.

The study project has been reviewed by the Institutional Review Board at University of Pittsburgh and meets all the necessary criteria for an exemption (IRB#: PRO13080310) in December 2013.

Our download of data from the ResearchGate Q&A platform utilized ResearchGate's "activity stream" design, which provides a list of all the latest user actions. During downloading, we used a newly created account in order to avoid any personalization bias, and then collected the list of question threads by repeatedly reaching the bottom of the webpage until the activity stream was completely expanded.

Figure 2 illustrates the sample collection process. After finishing a preliminary collection and analysis of LIS threads (with 413 records in 33 questions) in November 2013, we then expanded to two more disciplines "History or Art" (Art) and "Astrophysics" (Phy) in October 2014. Followed the data collection process described above, we collected 311 posts in 33 question threads for Art and 404 posts in 36 question threads in Phy.

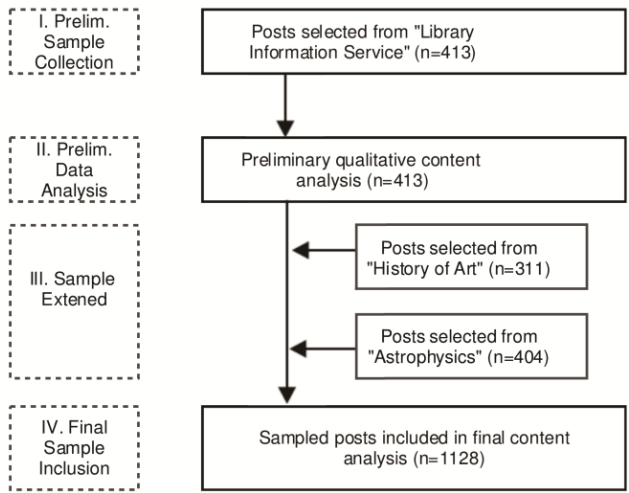

*Figure 2. Overview of data collection and coding scheme development*

For all responses to a given question, we captured each respondent's institution, the post date and time of the response, and the content of the response. In order to protect all scholars' anonymity, their profiles, publications, or RG scores were not identified or analyzed in the current study. On occasion a post included an email address for further collaboration, and this material was removed before saving the content to the dataset. Since we do not retrieve or store any information from users' profile, our data collection method does not violate the Terms and Conditions for ResearchGate, at Article 5-1, 6-2, and 6-3.

Combining the three disciplines, the overall sample consists of 1128 posts in 107 question threads. The samples in each discipline are comparable and sufficient in size to related previous works for a qualitative content analysis, which include Liu and Tsai (2008) with 408 notes and 140 messages, Pena-Shaff & Nicholls (2004) with 494 messages, and Harper, Weinberg, Logie, and Konstan (2009) with 300 questions.

The current dataset has two prior works. After the preliminary analysis in LIS, we reported on members' communication networks (Goodwin et al., 2014). The second work used a subset (1021 answer posts) and collected six more variables to evaluate the factors influencing user ratings (i.e., upvotes and downvotes) of answers on ResearchGate and predictive models of answer quality and was reported in Li et al. (2015).

**Coding Scheme Development**

Our preliminary coding scheme involved a two-level classification. We identified four overarching categories, with a binary coding method: questioner's intentions, detailed characteristics of the post, social cues, and consensus building. Under each top-level category, we further specified several sub-categories. The following is a detailed discussion of the themes and their characteristics.

### Questioner's intentions

Firstly, we focused on a topology that could reflect questioners' intentions. Drawing a classification from Fahy et al. (2001), we first categorized the whole 38 threads in LIS into information seeking questions (N=17), discussion seeking questions (N=17), and non-questions (N=4) in December 2013. We followed the same coding rule and classified *History of Art* and *Astrophysics* in October 2014.

Information seeking questions (hereafter: IQs) are those Fahy (2001) called a "vertical question", where a correct answer exists if the right authority or reference may be provided to support the answer. Some examples we found are:

> *"Can you recommend (empirical) studies on the data sharing behaviour of (academic) researchers? (LIS)"*
>
> *"Does anyone know about the sizes of remnant debris of comet Ison and do their trajectories intersect with earth (Phy)"*

Discussion seeking questions (DQs) are called "horizontal questions" by Fahy et al.'s definition (2001). There may not be a right answer for these questions, but instead, more responses are invited to help provide a plausible answer, or at least help shed more light on the question itself. Some observed examples are:

> *"In your opinion and experience which one is the better way that students could learn and enjoy History? (Art)"*
>
> *"Would wormholes be useful in flat cosmological space (Phy)"*

Non-questions (NQs) contain question threads that we could not classify into either IQs or DQs. Two examples we found of NQs are:

> *"Information Literacy: The Fourth R. (LIS)"*
> *"Web 2.0, 3.0 and web-based library services (LIS)"*

In the example above, the question initiator of "Information Literacy: The Fourth R." shared information without requesting further feedback; whereas the initiator of "Web 2.0, 3.0 and web-based library services" misunderstood how to use ResearchGate Q&A. The question initiator created a sub-topic under Library Information Services by simply posting key words, rather than explicitly providing any starting content or requesting any feedback.

### Content features of the post

In order to capture any additional content features of a post, based on classifications presented in (Pena-Shaff & Nicholls, 2004; Zhu, 1996), we captured five types of content features: CF1x, CF2, CF3x, CF4x, and CF5 in Table 1. These five sub-categories were applied to both questions and responses. Table 1 provides justifications for each coding category. All these characteristics have a prefix "self-provide" to indicate that the question initiator was prompted only by themself to provide the content as part of a question. A response does not include this "self-provide" prefix. Note that we made three modifications to the coding scheme for the final coding system: CF1x, CF3x, and CF4x.

*Table 1. Coding schemes and modifications*

| Top-level categories | Preliminary coding system | Final coding system | Justifications of final coding system "Apply this code as "1" if" |
|---|---|---|---|
| Questioner's intentions | QI1. Seeking information | -- | a question initiator asked a "vertical question," where a correct answer exists if the right authority or reference may be provided to support the answer. |
| | QI2. Seeking discussion | -- | a question initiator asked a "horizontal question," where more responses are invited to help provide a plausible answer. |
| | QI3. Non-questions | -- | coders could not classify a question into either IQs or DQs. |
| Content features of the post | CF1. Adding information | CF1x. Adding factual information | a post (either a question or a response) contains information based on facts. |
| | CF2. Providing Resources | -- | a post shares hyperlink, citations, files, research objects as sources. |
| | CF3. Referring to other researchers | CF3x. Referring to theories, famous concepts or frameworks in a discipline | a post contains information which related to famous concepts, scientific law, theatrical frameworks in a discipline without a citation. e.g., the Newton's second law, Ranganathan's Laws of Library Science. |
| | CF4. Providing opinions | CF4x. Providing opinions and feedback to others | a post provides opinions to the question and/or feedback to any other users in the thread. |
| | CF5. Providing personal experience | -- | a post contains a user's self-disclosed information such as their work background, life experience, or their attempts regarding a research-related decision. |
| Social cues | SC1. Comfort | -- | a post contains an emotional supportive message. |
| | SC2. Politeness | -- | a post contains short and warm greeting such as "thank you" and "all the best" |
| | SC3. Open for a further contact | -- | a post contains a responder's contact information such as an email address |
| Consensus Building | CB1. Agreement | -- | a post contains agreement or positive feedbacks about others' posts |
| | CB2. Disagreement | -- | a post contains disagreement or negative feedbacks about others' posts |

For the preliminary coding, we did not specify the definition of "information" in CF1. *adding information*, thus creating unclear adoptions among coders. We then modified CF1 to CF1x: *adding "factual" information*. Another modification was applied to CF3. At first, if a question initiator in LIS mentioned Ranganathan's laws without providing a direct link, we categorized this as *refer to other researchers* (CF3). However, when expanded to History of Art and Astrophysics, we observed that many answerers directly referred to famous theories or scientific laws. Therefore, we refined CF3 by providing a clearer and more sound description (CF3x in Table 1). We also refined CF4 by acquiring both *opinions to the question as well as feedback to other responses* (CF4x). After these modifications to these three categories' definitions and scope, we revisited all 413 posts in LIS and ensured the final coding scheme was adopted in all 1128 posts. The code CF5 described a code that contains elements of personal experience relating to the question. Coders detected this element when an RG user disclosed their work background, life experience, or described the attempts they have tried regarding a research-related decision. For example:

*"As a lay person with no formal training in physics, I wholeheartedly agree! (Phy252)"*

### Social cues

We detected social cues to capture affective and emotionally supportive messages behind academics' interaction. This category includes offering comfort to another user (SC1) - which helps identify emotionally supportive messages that encourage other academics - and politeness (SC2) e.g.,

*"First of all I want to note the pleasure I have to talk with this fai[t]hful academic assembly.. I send you my warm greeting from Paris. (Art182)."*

We were also interested in whether academics on RG extend their communication to an offline setting. Therefore, we also detected if a post included any contact information, such as an email address, for an offsite discussion (SC3).

### Consensus Building

A post coded in this category had to explicitly state an agreement or positive feedback (CB1) with an initiator or respondent through language such as "I agree" or "I think … is right." For the sub-category disagreement (CB2), coders detected the user explicitly providing negative feedback or disagreement to others through language such as "I disagree."

### Inter-coder Reliability

Table 2 summarizes the coding process and the inter-coder reliability for our preliminary coding and final coding results. The sample in Phase I was coded by two coders. The Cohen's kappa coefficients at the theme level and at the sub-categories level were .80 and .61, indicating the coding was reliable (Viera & Garrett, 2005).

*Table 2. Overview of coding process and inter-coder reliability*

| Phase | Sample | Sample size | Testing sample size | # of Coders | Top level categories | | sub-categories | |
|---|---|---|---|---|---|---|---|---|
| | | | | | Percentage | Cohen's kappa | Percentage | Cohen's kappa |
| I. | Library Information Service | 413 | 413 | 2 | 0.911 | 0.8 | 0.919 | 0.61 |
| II. | History of Art & Astrophysics | 715 | 60 | 3 | 0.904 | 0.77 | 0.936 | 0.59 |
| III. | | | 174 | 3 | 0.925 | 0.82 | 0.914 | 0.62 |

In Phase II, we set a minimum acceptable level for the preferred indices based on scholars' suggestions: a percent agreement of .85 or greater (Lombard et al., 2010), and Cohen's kappa coefficient of .5 or greater (Viera & Garrett, 2005).

As we created a codebook based on Phase I for the participating coders in Phase II, we decided to ensure the reliability of coding by assessing a small number of units. In the first pilot testing, three coders paired and examined 60 posts with an overall agreement rate of 90.38% and .77 (Cohen's kappa) at the top level; 93.56% agreement and .59 (Cohen's kappa) at the sub-category level. Based on coders' feedback, we refined the coding instructions (i.e.,

CF1x, CF3x, and CF4x in Table 1). After the pilot training, three coders divided all the posts into groups of 715 posts and coded each group individually.

Once all posts were coded, we then randomly drew a small sub-set of one-fourth posts (N=174) for ensuring inter-coder reliability. We re-assigned these 174 posts and ensured every post was coded by two coders. The Cohen's Kappa value was .82 at the theme level and .62 at the sub-category level, which suggests our coding process rigidly followed the literature, and the results on both levels are reliable.

<div align="center"><strong>RESULTS</strong></div>

## Overview

Among the 1128 posts collected, we could not locate the authors' information of 31 posts because their profiles were deactivated. Among the remaining 1097 posts, we found 478 unique users, and each of them contributed 2.3 (SD=3.75) posts on average. 43 users posted 5 times or more, whereas 312 users contributed only one post.

Among these 478 scholars, we were able to locate 391 distinct research institutions from 66 countries using a Geotag service (findlatitudeandlongitude.com), as visualized in Figure 3. The countries with the most scholars were USA (N=73, 18.7%), India (N=57, 14.6%), UK (N=37, 9.5%), and Germany (N=25, 6.4%). The geographic distribution is largely consistent with Thelwall and Kousha (2015)'s. When narrowing down to individual disciplines (Figure 4, 5, and 6), we observed that the U.S., India, and European countries remain the top contributors for each discipline. While India is a top contributor overall, it accounts for a much smaller portion in Art or LIS.

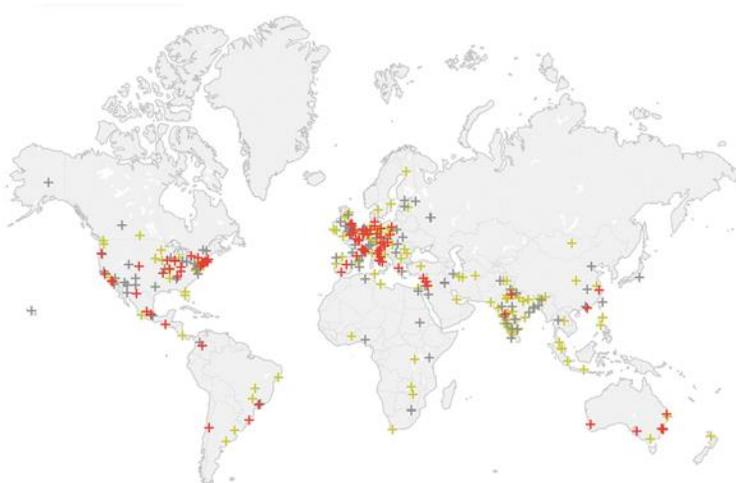

*Figure 3. Geographic distribution of sampled users in three disciplines (unit: scholarly institution)*

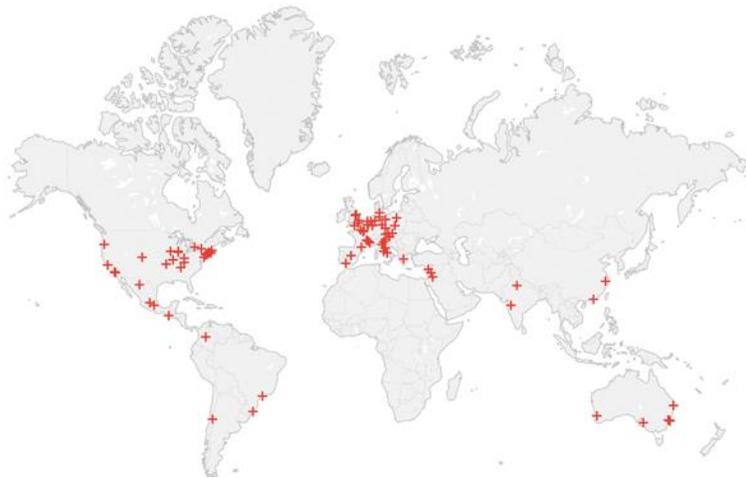

*Figure 4. Geographic distribution of sampled users in History of Art (unit: scholarly institution)*

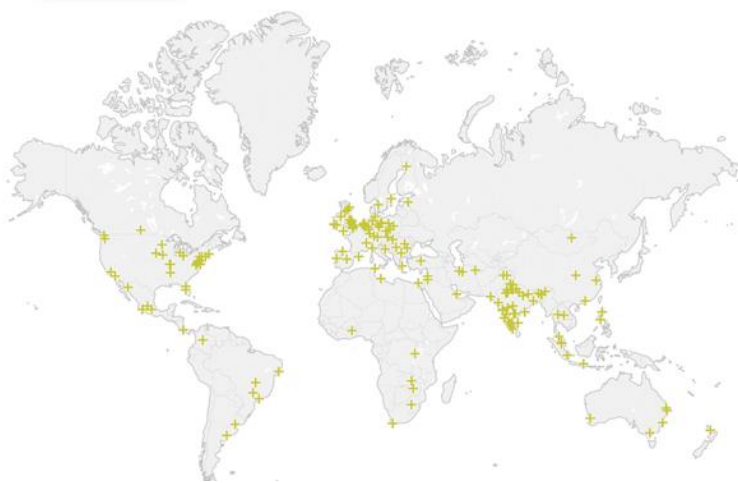

*Figure 5. Geographic distribution of sampled users in Library Information Sciences (unit: scholarly institution)*

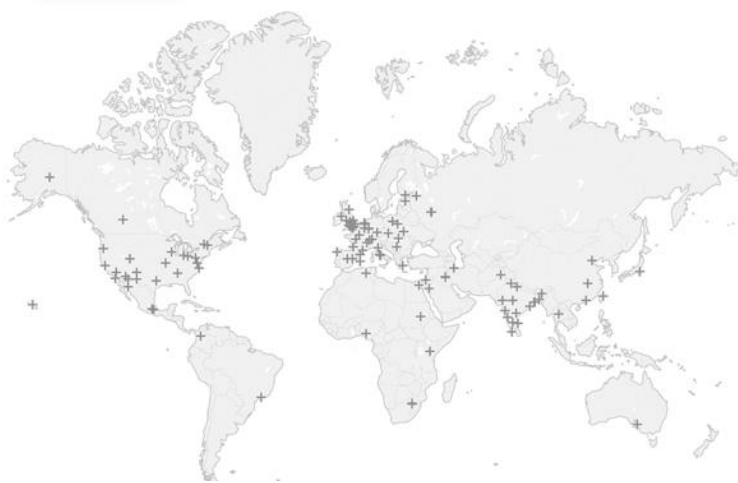

*Figure 6. Geographic distribution of sampled users in Astrophysics (unit: scholarly institution)*

On average, a question received 10.54 responses (SD=13.5, Mdn= 7.0). The median of the response time for the first response was 15.36 hours, whereas the median of the time interval between each response was 7.9 hours.

**Question Characteristics**

The results of content analysis to study the distribution of questioners' intentions for all 107 questions are summarized in Table 3. We found that Art has more discussion-seeking threads, whereas Phy has more information-seeking questions. After removing instances of the non-question, a chi-square test suggested that the distribution of information or discussion threads did not differ by discipline, $\chi^2$ (2, N = 103) = 2.66, p = .264.

*Table 3. Sampled questions, by discipline and quesitoners' intentions*

| Discipline | Seeking information | | Seeking discussion | | Non-questions | | TOTAL |
|---|---|---|---|---|---|---|---|
| | N | % | N | % | N | % | |
| **History of Art (Art)** | 12 | 36.4% | 21 | 63.6% | 0 | 0% | 33 |
| **Library Information Service (LIS)** | 17 | 44.7% | 17 | 44.7% | 4 | 4.6% | 38 |
| **Astrophysics (Phy)** | 20 | 55.6% | 16 | 44.6% | 0 | 0% | 36 |
| **TOTAL** | 49 | 45.8% | 54 | 50.5% | 4 | 3.7% | 107 |

Figure 7 shows the distribution of measures between factors of disciplines and factors of question types. The number of total responses (Figure 7a), hours to the first response (Figure 7b), and length of the question (Figure 7c) were subjected to two-way analysis of variance. However, no effect was statistically significant at the .05 level, suggesting there is no sufficient evidence to conclude the number of total responses, time to first response, or question length vary because of the discipline, question intention, or their factor interactions.

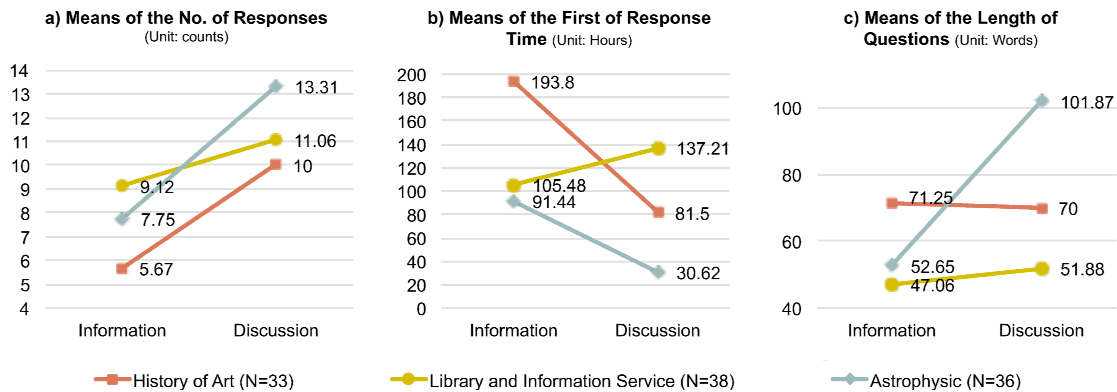

In Table 4, we observed some similarities among the disciplines. Twenty-nine of 107 question initiators (27%) provided resources such as documents or URLs in the initial question. For example, one question initiator in LIS introduced OCLC's report "U.S. Library Consortia: Priorities & Perspectives" in order to discuss the future of librarianship.

*Table 4. Question characteristics **in content feature***

| | History of Art (N=33) | | Library information services (N=38) | | Astrophysics (N=36) | | Total (N=107) |
|---|---|---|---|---|---|---|---|
| | N | % | N | % | N | % | N |
| Specifically request resources | 11 | 33.3 | 3 | 7.9 | 0 | 0 | 14 |
| CF1. Self-adding factual information | 13 | 39.4 | 3 | 7.9 | 18 | 50 | 34 |
| CF2. Self-provide resources | 13 | 39.4 | 4 | 10.5 | 12 | 33.3 | 29 |
| CF3x. Self-referring to theories, famous concepts or frameworks in a discipline | 7 | 21.2 | 3 | 7.9 | 12 | 33.3 | 22 |
| CF4x. Self-providing opinions and feedback to others | 6 | 18.2 | 1 | 2.6 | 3 | 8.3 | 10 |
| CF.5 Self-providing personal experience | 9 | 27.3 | 0 | 0 | 4 | 11.1 | 13 |

Note: The coding was not mutually exclusive, thus the overall characteristics can exceed 100%.

We also observed some discipline-specific characteristics of the questions. For example, we found 11 out of 33 History of Art questioners requested specific resources, whereas we found only few instances in Phy (N=0) and LIS (N=3).

As another example of discipline-specific characteristics, half of Phy questioners and more than one-third of Art questioners provided factual information in order to better describe their questions. For example, in the thread "Can electric and magnetic forces be viewed as space curvature with particular limitations like gravity?" the questioner first described "Long back perception on gravity was changed from a kind of force to a phenomenon which actually bends space-time in its influence (Phy091)."

## Response Characteristics

Figure 8 shows the distribution of responses between factors of disciplines and question types. For the length of responses, a two-way analysis of variance yielded a main effect for the responses' disciplines, $F_{(2, 983)} = 7.26$, $p = .001$, such that the average length of a response was significantly higher for Astrophysics (M = 100.17) than for History of Art (M = 88.70) and LIS (71.03). The main effect of questioners' intentions was also significant, $F_{(1, 983)} = 7.58$, $p=.006$. The interaction effect was non-significant, $F_{(2, 983)} = 1.21$, $p=.3$. These tests suggest that a response to a discussion-seeking question is more likely to contain richer and longer content than a post in information-typed threads, no matter the discipline.

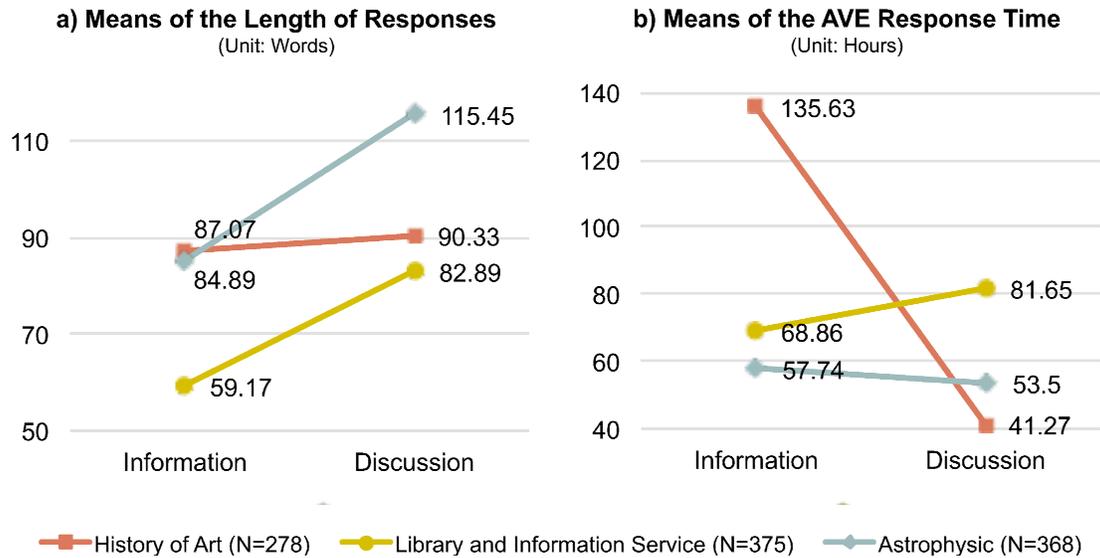

**a) Means of the Length of Responses**
(Unit: Words)

**b) Means of the AVE Response Time**
(Unit: Hours)

■ History of Art (N=278)  ● Library and Information Service (N=375)  ◆ Astrophysic (N=368)

*Figure 8. Measures of Responses*

When we examined the factors that were associated with response time, the main effect of questioners' intention yielded an F ratio of $F(1, 887) = 8.783$, p=.003, indicating that the response time was significantly longer for posts in information-typed threads (M = 87.41) than for discussion messages (M = 58.81). The main effect of disciplines was non-significant: $F(2, 887) = 3.88$, p=.021, greater than .01 level. However, the interaction effect was significant between these two factors: $F(2, 887) = 9.9$, p < .001, indicating that the questioner intention effect was greater in the discipline of History of Art than in the other disciplines. The tests suggest that, compared with Library Information Services and Astrophysics, the discussion posts in the History of Art discipline tend to receive responses within a shorter time.

Table 5 presents the distribution of answers' characteristics across three disciplines. After a Bonferroni correction which adjusted the alpha to be at .005, a Chi-square test suggested the distribution of instances among the three disciplines were significantly different. Respondents in Astrophysics were more likely to provide factual information ($\chi^2$ (2, N = 1021) = 69.941, p < .0001. Cramer's V= .262) and refer to others ($\chi^2$ = 61.00, p < .0001. Cramer's V= .245), whereas in History of Art we found more instances of providing resources ($\chi^2$ = 53.461, p < .0001. Cramer's V= .229) and personal experience ($\chi^2$ = 12.60, p = .002. Cramer's V= .111). Agreement and disagreement were also found to be significant, with less agreement in LIS and more agreement in Astrophysics.

*Table 5. Response characteristics*

| | History of Art (N=278) | | Library information services (N=375) | | Astrophysics (N=368) | | Total (N=1021) | |
|---|---|---|---|---|---|---|---|---|
| | instances | % | instances | % | instances | % | N | % |
| CF1. adding factual information ** | 110 | 39.6% | 100 | 26.7% | 209 | 56.8% | 419 | 41.0% |
| CF2. provide resources ** | 130 | 46.8% | 75 | 20.0% | 112 | 30.4% | 317 | 31.0% |
| CF3x. referring to theories, famous concepts or frameworks in a discipline ** | 34 | 12.2% | 15 | 4.0% | 86 | 23.4% | 135 | 13.2% |
| CF4x. providing opinions and feedback to others | 139 | 50.0% | 190 | 50.7% | 215 | 58.4% | 544 | 53.3% |
| CF.5 providing personal experience * | 46 | 16.5% | 38 | 10.1% | 29 | 7.9% | 113 | 11.1% |
| SC1. Comfort # | 2 | 0.7% | 1 | 0.3% | 1 | 0.3% | 4 | 0.4% |
| SC2. Politeness | 90 | 32.4% | 118 | 31.5% | 89 | 24.2% | 297 | 29.1% |
| SC3. Open for a further contact # | 0 | 0.0% | 8 | 2.1% | 1 | 0.3% | 9 | 0.9% |
| CB1. Agreement ** | 64 | 23.0% | 34 | 9.1% | 64 | 17.4% | 162 | 15.9% |
| CB2. Disagreement** | 20 | 7.1% | 12 | 3.2% | 55 | 14.9% | 131 | 12.8% |

Note: N=1021; *: p<.005; **: p<.0001; #Excluded by the Chi-square test because the sample size is insufficient.

There was no significant difference observed in providing personal opinions or politeness: this finding suggests that these behaviors can be relatively common across disciplines.

**Resources**

 As shown in Figure 9, we further analyzed the types of resources that scholars provided to the questions. We created a sub-set of 317 posts, based on the posts that we identified with CF2. *provide resources*. We recognized 13 mutually exclusive categories of provided resources, including:

- Expert resource: providing contacts of expert (i.e., a name),
- Traditional academic publications: citations, links to a book information, and links to paper files
- News: links to news articles or magazine articles
- Digital objects, especially on social media: links to projects, grants, images (links or directly uploaded), links to software information, videos (e.g., a YouTube link), links to Wikipedia entries, blogs, and other answered RG questions.

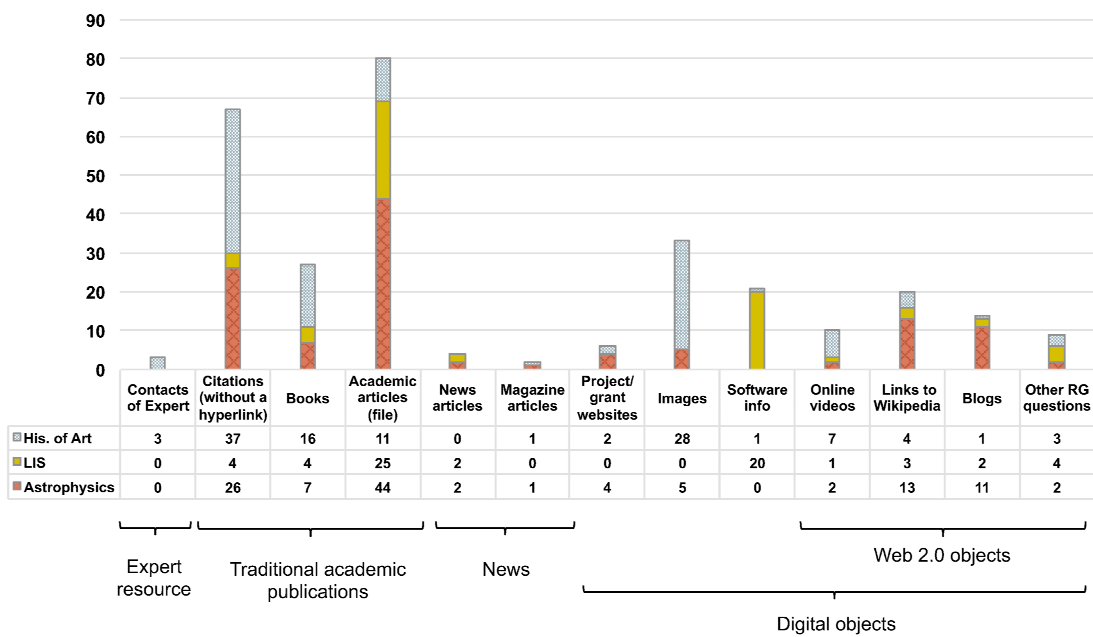

| | Contacts of Expert | Citations (without a hyperlink) | Books | Academic articles (file) | News articles | Magazine articles | Project/ grant websites | Images | Software info | Online videos | Links to Wikipedia | Blogs | Other RG questions |
|---|---|---|---|---|---|---|---|---|---|---|---|---|---|
| His. of Art | 3 | 37 | 16 | 11 | 0 | 1 | 2 | 28 | 1 | 7 | 4 | 1 | 3 |
| LIS | 0 | 4 | 4 | 25 | 2 | 0 | 0 | 0 | 20 | 1 | 3 | 2 | 4 |
| Astrophysics | 0 | 26 | 7 | 44 | 2 | 1 | 4 | 5 | 0 | 2 | 13 | 11 | 2 |

Expert resource | Traditional academic publications | News | Web 2.0 objects

Digital objects

*Figure 9. Types of resources provided*

For the 130 posts in History of Art which provided resources, citations (N=37, 28.5%), images (N=28, 21.5%), and books (N=16, 12.3%) were the most common categories. In 75 instances in LIS, two major categories emerged: academic articles (N=25, 33.3%) and software information (N=20, 26.7%). As for the 112 instances in Astrophysics, overwhelmingly, we found answerers were most likely to point out academic papers rather than books or other resources. They either uploaded a paper to the community (N=44, 39.3%) or mentioned it through a citation (N=26, 23.2%).

Overall, we noticed that the traditional academic resources such as citations and books still play a very important role for providing resources. Multimedia, such as online videos and images, also take a dominant position in History of Art. We also observed that social media resources such as Wikipedia, blogs articles, and even other ResearchGate Q&A threads have been provided to academic peers.

## DISCUSSION

In this section, we discuss insights gained from the research findings.

### Insights on Questioners' Intentions

For our first research question, we observed that users in the three disciplines were not significantly likely to diverge in asking information- or discussion-seeking questions and that the type of question asked had no effect on characteristics of the responses.

Furthermore, within information-seeking questions, we observed that some are derived from individual's information needs and aimed to satisfy that personal request, whereas many other questions are related to community's information needs. These kind of questions might not be easily spotted by content since on an ASNS it is often dependent on how the virtual community receives the question. The initial questioner in a thread asked for software to check plagiarism, and the Library and Information Services community responded overwhelmingly with 55 responses, many of which were directed to other respondents rather than the question initiator in order to explain a complex ethical dilemma. Another thread in LIS asked for a list of social networks, and the community did not respond as enthusiastically, with only 8 responses, showing the direction of the community's unspoken information needs. This illustrates the complexity of questioners' intentions and the dynamics of ASNS online communities.

## Discipline Influence on Scholars' Sharing Behaviors on ASNSs

Our study identified some similarities among scholar's ASNS behaviors in the three disciplines. On the question side, we found that discussion-seeking questions received more responses than information-seeking ones in all disciplines, and there was no significant difference in terms of the time to first response or length of the questions. Then, on the response side, there was no significant difference in social elements or providing personal opinions among the different disciplines either, which indicates that the sharing behaviors in these disciplines were similar when expressing social elements.

Our second research question examined the characteristics of the responses on ResearchGate Q&A. We found that History of Art, our example in humanities, took a longer time to answer information-seeking questions than discussion-seeking ones. The questioners in History of Art also tended to ask for resources directly, such as specifically asking for a press or picture. In the literature, such direct demands from group leaders were found to be associated with the positive development of the community (Jeng, He, Jiang, & Zhang, 2012). At the same time, we found a wide variety of resources that History of Art users requested and provided, including the largest number of books among the three disciplines. This is consistent with Brockman et al. (2001), who found that scholarly materials used in humanities research are drawn from "a wide variety of types of resources" and scholars in humanities "rely on books more heavily than on journals"(p.10).

Our example discipline in social science, LIS, exhibited some different behavior patterns. For instance, it took a longer time for the LIS users to answer discussion-typed questions than information-typed questions, which is different than the other two disciplines. Furthermore, responses in LIS rarely referred to theories or famous concepts. One reason could be that LIS professionals and scholars are trained to provide citations rather than direct answers. Another reason could be that the LIS is a discipline with a strong practical emphasis on services, which makes LIS users value practical experience more. As for preferred resources, we found that our results in LIS are consistent with Ellis' information-seeking model for social scientists (Ellis, Cox, & Hall, 1993), in which books, journal articles, and newspaper articles are important resources.

Astrophysics, our example in the natural sciences, had the most responses that "refer to theories, famous concepts or frameworks" among the three disciplines, and more than a half of responses in it contained factual information. These results are consistent with the literature (e.g., Holland and Powell, as cited in Case, 2012) reporting that scientists and engineers like to ask and confirm factual or ready information with their colleagues, especially using "word of mouth" when they seek information.

## Comparison of Behaviors on ASNSs and Other Social Platforms

Before the current form of ASNSs like ResearchGate, traditional listservs had been the only popular platforms for scholars to express their opinions and exchange information. Previous research found that the rich content expressed on these earlier social platforms still exhibited the style of scholarly writing (Gu & Widén-Wulff, 2011). We obtained the same impression from the sample posts in all three disciplines collected from ResearchGate Q&A. The posts adhered to some strict formats that only exist in academic writing, such as referring to theories or frameworks, or providing resources.

However, contrary to Veletsianos' (2012) conclusion that social media sites such as Twitter are useful for promoting scholars' research activities, conference trips, or recommending information, we did not witness that users on ResearchGate have similar broadcasting patterns. One possible explanation could be that since ResearchGate's interface has changed, scholars in the Q&A style interface are more likely to reply to individual posts, rather than broadcast to the entire community or other potential readers (Goodwin, Jeng, & He, 2014).

## Managerial Implications to ASNS Design

Based on the results of our study, we propose the following suggestions for continued evolution of scholarly Q&A platforms for supporting scholars' information exchange.

*The system should consider both one-to-one engagements and broad interactions.* We have observed many times in our study that a scholar may want to explicitly direct information to a single scholar in some posts, but may want to disseminate other discussions to a broad group of scholars in other exchanges. We think ASNS Q&A can borrow the pointed interaction functions (e.g., "#", "@", "+" annotations) that are popular in general SNS platforms such as Twitter and Google+ to enhance this kind of interaction. Another benefit is these existing annotations would require minimal learning efforts from users since they are so popular in generic social sites.

*The support should minimize interference and confusion.* One important limitation in the current design of ResearchGate Q&A is that there is no indication from the question initiator to mark whether or not a question has been satisfactorily answered. On the one hand, we understand that many academic questions do not have a definite answer, thus do not have clear closure. On the other hand, it is difficult for scholars who have limited time and resources to prioritize the set of questions that are both interesting and open.

*The threads should have longer life span.* Unlike some popular leisure topics, many academic problems and topics, even after some time has passed, are still relevant and deserve further attention. Currently ResearchGate promotes the latest questions, but lacks effective mechanisms to engage in older questions. We suggest that some form of question rotation scheme (e.g., Wikipedia's today's featured article) may be useful to ensure scholars' attentions to older questions to address the longer life span of scholarly information.

## CONCLUSION

Using ResearchGate Q&A platform as the focus of our case study, this paper presents an evaluation of whether and to what extent existing ASNS platforms can facilitate scholarly information exchange. We conducted content analysis to examine 1128 posts collected from ResearchGate Q&A. These posts cover three different disciplines: Library and Information Services (as a case of social sciences), History of Art (humanities), and Astrophysics (science).

Our results show that there are similarities and differences among scholars' information exchange behaviors in the three disciplines. In particular, we found that scholars seek information and discussion equally among the three disciplines and frame their questions with external resources with the same regularity. Unique to History of Art Q&A, questioners were more likely to request a known resource. LIS questioners were more likely to not frame their information need with supplemental information. In posting answers, scholars took a longer time to respond to their peers in information-typed questions than in discussion ones, and users in Astrophysics were more likely to provide factual information and write longer responses. We also concluded that providing personal opinions and being polite are common behaviors across all three disciplines. Based on these results, we discussed several implications to the study of scholars' behaviors on ASNSs, as well as suggestions for ASNS design. We argue that an ASNS should stimulate scholarly interactions by minimizing confusion, helping improve the clarity of questions, and promoting content management.

In the future, we would like to extend our framework along two dimensions to examine its generalizability and to make it more robust. First, we plan to apply our coding scheme to other disciplinary groups on ASNSs, such as engineering related fields in applied sciences or health science domains. Second, we plan to compare the content and user behaviors on ASNSs with those in different scholarly information exchange scenarios, such as a face-to-face interaction or an open peer review platform. Another promising future direction is to implement the design suggestions proposed in the previous sections and evaluate the applications in a realistic setting.

## ACKNOWLEDGEMENT

This study is partially supported by the National Natural Science Foundation of China's International Joint Research Project "Research on Knowledge Organization and Service Innovation in the Big Data Environments" (Project#71173249).